\documentclass[aip,numerical,floatfix,groupedaddress,reprint]{revtex4-1}
\usepackage[utf8]{inputenc}  
\usepackage[T1]{fontenc}  
\usepackage[english]{babel} 
\usepackage{epsfig} 
\usepackage{parskip}
\usepackage{setspace}
\usepackage{caption3} 
\DeclareCaptionOption{parskip}[]{} 
\usepackage[small]{caption}
\usepackage{subcaption}
\usepackage{subfig}
\usepackage{ifpdf}
\usepackage{nomencl}
\usepackage{empheq}
\usepackage{nicefrac}
\DeclareMathSizes{10}{10}{10}{10}
\addto\captionsenglish{}
\begin{document}
\title{Radiative thermal rectification using superconducting materials}
\author{Elyes Nefzaoui}
\email[E-mail : ]{elyes.nefzaoui@univ-poitiers.fr}
\author{Karl Joulain}
\email[E-mail : ]{karl.joulain@univ-poitiers.fr}
\author{Jérémie Drevillon} 
\author{Younès Ezzahri}
\affiliation{\small Institut Pprime, CNRS-Universit\'e de Poitiers-ENSMA, D\'epartement Fluides, Thermique, Combustion, ENSIP-B\^atiment de m\'ecanique, 2, Rue Pierre Brousse, F 86022 Poitiers, Cedex, France}
\date{\today}
\begin{abstract}
Thermal rectification phenomenon is a manifestation of an asymmetry in the heat flux when the temperature difference between two interacting thermal reservoirs is reversed. In this letter, we present a far-field radiative thermal rectifier based on high temperature superconducting materials with a rectification ratio up to $80\%$. This value is among the highest reported in literature. Two configurations are examined : a superconductor (Tl$_2$Ba$_2$CaCu$_2$O$_8$) exchanging heat with 1) a black body and 2) another superconductor, YBa$_2$Cu$_3$O$_7$ in this case. The first configuration shows a higher maximal rectification ratio. Besides, we show that the two superconductors rectifier exhibits different rectification regimes depending on the choice of the reference temperature, \textit{i.e.} the temperature of the thermostat. Presented results might be useful for energy conversion devices, efficient cryogenic radiative insulators engineering and thermal logical circuits development. 
\end{abstract}
\maketitle
Thermal rectification can be defined as an asymmetry in the heat flux when the temperature difference between two interacting thermal reservoirs is reversed. 
For a system in an initial state characterized by a given thermal gradient, the rectification ratio can be defined by :
\begin{eqnarray}
R & = & \frac{q_{FB} - q_{RB}}{max(q_{FB},q_{RB})}
\label{eq:Recitification}
\end{eqnarray}
where $q_{FB}$ is the heat flux in the initial state and $q_{RB}$ the heat flux when the thermal gradient is reversed. A non zero rectification means that a reversal of the thermal gradient induces, in addition to the reversal of the heat flux direction, a variation of its magnitude.
The realization of a device exhibiting such an uncommon behavior, a thermal rectifier for instance, would pave the way to the development of thermal circuits in the manner non-linear electronic devices marked the genesis of modern electronics\cite{Wang2008}. Indeed, as in electronics, thermal logical circuits need a thermal diode which can be defined as an ideal rectifier, \textit{i.e.} a one-way heat transmitter.
Consequently, an increasing interest has been given to thermal rectifiers during the past decade. 

First works focused on heat conduction devices and have led to rectification models based on non-linear lattices \cite{Terraneo2002,Li2004,Li2005,Hu2006}, graphene nano-ribbons \cite{Hu2009,Yang2009} and several other interesting mechanisms \cite{Segal2008,Yang2007}. Some authors went beyond the thermal rectification issue and proposed theoretical models of thermal logical gates \cite{Wang2007} and a thermal transistor \cite{Lo2008}.
Experimental implementations of rectifiers based on carbon and boron nitride nanotubes \cite{Chang2006}, semiconductor quantum dots \cite{Scheibner2008} and bulk cobalt oxides\cite{Kobayashi2009} have also been realized. Extensive reviews treating thermal rectification in solids can be found in Refs. \cite{Li2012,Roberts2011}.

During the last four years, a few authors tackled the question of radiation based thermal rectification.
A theoretical study and an experimental suggestion of a radiative thermal rectifier based on non-linear solid-state quantum circuits operating at very low temperatures (a few mK) have been first presented \cite{Ruokola2009}. A rectification ratio up to $10\%$ is predicted. Moreover, two theoretical schemes of radiative thermal rectification based on near-field thermal radiation control have lately been proposed \cite{Otey2010,basu2011}. A rectification ratio up to $30\%$ (according to the present paper rectification definition and using the references data) is theoretically predicted for temperature differences ranging between $100$ K and $300$ K. Comparable rectification ratios have also been reached, for the same temperature differences, by a Fabry-Pérot cavities based far field radiative thermal rectifier we recently presented \cite{Nefzaoui2013}. We reported a maximal rectification ratio of $19\%$. Shall we note here that the interacting bodies in all these early radiative rectifiers are discrete modes resonators : bulk materials supporting surface resonances, surface phonon-polaritons \cite{Otey2010} or suface plasmon-polaritons \cite{basu2011} for instance, nanostructured materials with cavity modes \cite{Nefzaoui2013} or nonlinear quantum resonators \cite{Ruokola2009}. Rectification is therefore achieved by controlling the coupling between the two bodies modes. Switching from a state where the two thermal reservoirs modes are strongly coupled, forward bias (FB), to a state of weak coupling, reverse bias (RB),  leads to a decrease in the exchanged radiative heat flux, thus to a thermal rectification.

More recently, the idea of broadband radiative thermal rectification (BRTR) emerged with experimental and theoretical studies of radiative heat flux modulation with a phase change material (PCM), $VO_2$ for instance \cite{Zwol2011a, Zwol2011b, Zwol2012a, Zwol2012b}. Heat flux contrasts up to $80\%$ and $90\%$ have been experimentally proven in the far-field \cite{Zwol2012a} and the near-field\cite{Zwol2012b}, respectively. Theoretically, a contrast of near-field radiative heat flux of almost $100\%$ is predicted for a bulk plane-plane glass/VO2 configuration \cite{Zwol2011a}. Shall we here emphasize that these contrasts are only observed around $VO_2$ metal-insulator transition temperature ($67^{\circ}$C) which strongly restricts their potential practical scope. Exploiting these contrasts to design a $VO_2$-based thermal rectifier has then been proposed \cite{Huang2013,Ben2013diode} and rectification ratios up to $70\%$ and $90\%$ for small and large temperature differences are reported, respectively.

Consequently, BRTR seems to be a promising path for efficient thermal rectification. In principle, a BRTR can be simply achieved with two planes of different random materials separated by a vacuum gap. Indeed, since non-identical materials are not likely to have identical thermo-optical properties (TOP), \textit{i.e.} the same temperature dependence of their optical properties, inverting the temperature gradient between the two planes would modify the spectral distribution of the exchanged radiation, and very probably the net heat flux. However, randomly chosen materials wouldn't lead to good rectifiers since TOP are generally very small\cite{Touloukian1970,Touloukian1972}, \textit{i.e.} optical properties vary slowly with temperature. Hence, PCM would be a relevant alternative since their optical properties exhibit strong variations with respect to temperature around the transition temperature $T_c$. 

Superconducting materials are a subset of PCM which has been thoroughly studied during the past century. In this letter, we propose a far-field radiative thermal rectifier based on high temperature superconductors. These materials are considered for essentially two reasons. First, a general feature of these materials TOP when they are switched from the normal to the superconducting state is that their reflectance significantly increases to almost $1$ in the far infrared while the mid-infrared spectrum remains almost unchanged \cite{Ginsberg1992}. Indeed, the dielectric function of these materials can be modeled by a sum of a Drude model and a finite number of Lorentz oscillators. Drude model accounts for the contribution of free carriers which are responsible for far infrared features. They are strongly affected by the phase transition while Lorentz oscillators account for the contribution of bound charges which are not affected by the phase change. On the other hand, a black body at $T \simeq 100$ K (a typical transition temperature for a high temperature superconductor) emits almost $90\%$ of its intensity in the spectral range above  $20$ $\mu$m, \textit{i.e.} in the far infrared. This is exactly the spectral domain where these materials show large TOP which makes them good candidates for radiation based thermal rectifiers. Two configurations are investigated : 1) a single superconductor device : a black body exchanging with Tl$_2$Ba$_2$CaCu$_2$O$_8$ ($T_{c,1} = 125$ K) and 2) a double superconductor device YBa$_2$Cu$_3$O$_7$ ($T_{c,2} = 93$ K) exchanging with Tl$_2$Ba$_2$CaCu$_2$O$_8$. A rectification ratio of almost $80\%$ can be reached with the proposed implementation for small temperature differences around $T_{c,1} = 125$ K. This implementation can be generalized to operate in a large temperature range ($[0,150]$K) at comparable rectification levels since several high temperature superconductors, with different and tunable critical temperatures, can be used.

\begin{figure}[ht]
\begin{center}
\includegraphics[width=0.45\textwidth]{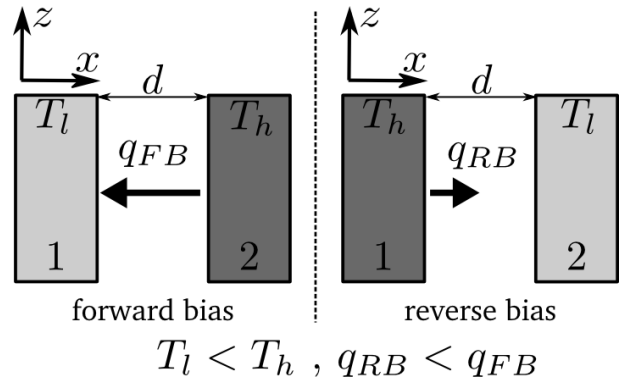}
\caption[]{Two parallel planar bodies separated by a vacuum gap of thickness $d$. In FB configurathion, $T_1 = T_l$ and $T_2 = T_h$. In RB, $T_1 = T_h$ and $T_2 = T_l$.}
\label{fig:PlanePlane}
\end{center}
\end{figure}

The figured out device is composed of two opaque thermal baths $1$ and $2$ at temperatures $T_1$ and $T_2$ respectively and exchanging heat through thermal radiation. 
Figure \ref{fig:PlanePlane} presents a schematic of the proposed device composed of two parallel planar bodies $1$ and $2$ separated by a gap of thickness $d$ and characterized by their optical properties, their emissivities for instance, and temperatures ($\varepsilon_1(T_1),T_1$) and ($\varepsilon_2(T_2),T_2$), respectively.
In FB, $T_1 = T_l$ and $T_2 = T_h$(Subscripts $l$ and $h$ stand for low and high temperatures, respectively). The two bodies temperatures are swapped in reverse bias.

The two bodies are assumed to be in vacuum so that $q$ is a radiative heat flux (RHF) density. The considered bodies radiative properties, in particular their emissivities and reflectivities ($\varepsilon$ and $\rho$ respectively), are completely governed by  their dielectric functions and geometries. In the case of opaque bodies, energy conservation and Kirchhoff's laws combination leads to the following relation between the monochromatic emissivity and reflectivity at a given temperature:
\small
\begin{eqnarray}
\varepsilon (T,\lambda) & = & 1 - \rho(T,\lambda)
\label{eq:EnergyConservation}
\end{eqnarray}
\normalsize
We also assume the two bodies are lambertian sources. $\varepsilon$ and $\rho$ are thus direction-independent. \\
The gap width $d$ is assumed to be much larger than the dominant thermal radiation wavelength (Wien wavelength) $\lambda_W(T) \simeq hc / 5 k_B T$ where $h$, $c$, $k_B$ and $T$ are Planck constant, the speed of light in vacuum, Boltzmann constant and absolute temperature, respectively. The net RHF density exchanged by the two media resumes then to the far field contribution which can be written \cite{Modest1993}:
\small
\begin{eqnarray}
q (T_1,T_2) & = & \pi \int_{\lambda=0}^{\infty} [I^0(\lambda,T_1)-I^0(\lambda,T_2)] 
\tau(\lambda,T_1,T_2) d \lambda
\label{eq:FluxDensity}
\end{eqnarray}
\normalsize
where 
\small
\begin{eqnarray}
I^0(\lambda,T) & = & \frac{2 h c^2}{\lambda^5}\frac{1}{e^{\nicefrac{h c}{\lambda k_B T}}-1}
\label{eq:Chap5-PlanckBBIntensity}
\end{eqnarray}
\normalsize
is the black body intensity at a temperature $T$ and
\small
\begin{eqnarray}
\tau(\lambda,T_1,T_2) & = & \frac{\left[1-\rho_1 (\lambda, T_1)\right]\left[ 1-\rho_2 (\lambda, T_2) \right]}{1-\rho_1 (\lambda, T_1) \rho_2 (\lambda, T_2)}
\label{eq:TransmissionCoeff}
\end{eqnarray}
\normalsize
is the monochromatic RHF density transmission coefficient between $1$ and $2$.

Now, we examine two configurations of Fig. \ref{fig:PlanePlane} device : 1) a configuration with a black body and a high temperature superconductor and 2) a configuration with two high temperature superconductors. An extensive review of different high temperature superconductors optical properties in the normal and superconducting state can be found in Ref. \cite{Ginsberg1992}.

\begin{figure}[ht]
\begin{center}
\includegraphics[width=0.45\textwidth]{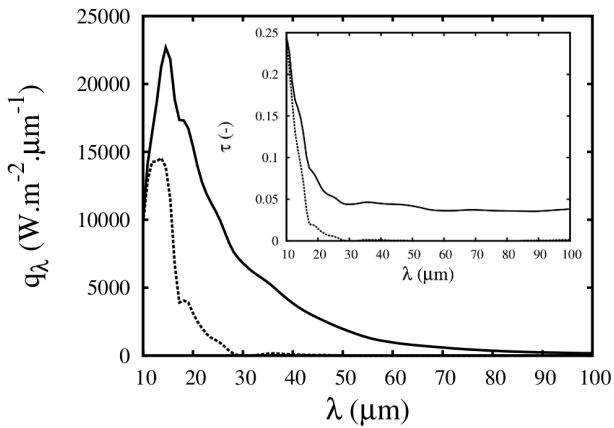}
\caption[]{Exchanged spectral radiative heat flux density given by the integrand of Eq. \ref{eq:FluxDensity} in FB (solid line) and RB (dashed line). Inset : spectral transmission coefficient given by Eq. \ref{eq:TransmissionCoeff} in FB (solid line) and RB (dashed line).}
\label{fig:TauPhiDevSingle}
\end{center}
\end{figure}

\begin{figure}[ht]
\begin{center}
\includegraphics[width=0.45\textwidth]{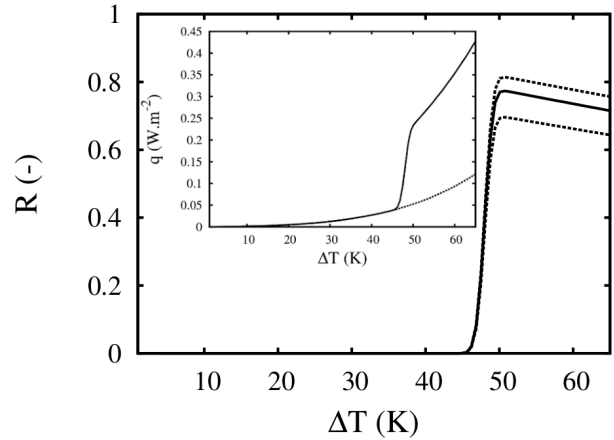}
\caption[]{Rectification ratio as a function of the temperature difference with $10\%$ uncertainty on the superconducting material reflectance experimental data (dashed lines). Inset : exchanged radiative heat flux density in FB (solid line) and RB (dashed line) as a function of the temperature difference.}
\label{fig:RvsT}
\end{center}
\end{figure}

Consider the case $T_l = 77$ K (liquid nitrogen boiling point) and $T_h = 140$ K. 
Body $1$ is a black body so that $\varepsilon (\lambda,T) = 1$.
Body $2$ is made of Tl$_2$Ba$_2$CaCu$_2$O$_8$, a high temperature superconductor with $T_c \simeq 125$ K.
Its far infrared dielectric function in the superconducting state shows a non negligible contrast with the normal state in the spectral range $[100,500]$ cm$^{-1}$, \textit{i.e.} $[20,100]$ $\mu$m.
Indeed, it's reflectance goes from $0.95$ in the normal state to almost $1$ in the superconducting state. 
We assume the optical properties in both the normal and the superconducting states temperature-independent which is more likely to be the case since we consider small temperature variations. The only significant variation of optical properties therefore occurs at the phase transition. Reflectance data of Tl$_2$Ba$_2$CaCu$_2$O$_8$ in the normal and superconducting states used in our calculations have been extracted from the corresponding plots given in Ref. \cite{Ginsberg1992}. 
Figure \ref{fig:TauPhiDevSingle} presents the exchanged radiative heat flux spectral density in forward and reverse bias respectively. I
A rectification ratio $R = 0.7$ is obtained. The behavior of the rectification ratio as a function of the temperature difference between the two bodies is presented in Fig. \ref{fig:RvsT}. The low temperature is fixed at $T_l = 77$ K and $T_h$ is given by $T_h = T_l + \Delta T$ with $\Delta T$ ranging over $[1,65]$ K.
No rectification is observed for $T_h < T_c$. The rectification ratio reaches a maximum $R_{s,max} = 0.79$ when $T_h$ is slightly larger than $T_c$ then slowly decreases. Since our calculations are based on experimental reflectance data, we also present in Fig. \ref{fig:RvsT} the uncertainty range of the rectification ratio (dashed lines) corresponding to a $10\%$ uncertainty on used reflectance data. This shows a relative uncertainty on the maximal rectification ratio ranging over $[-10\%,+5\%]$.

The inset of figure \ref{fig:RvsT} presents the exchanged RHF in FB (solid line) and RB (dashed line). Both fluxes increase over the whole range as $\Delta T$ increases. When $T_h < T_c$, $q_{FB} = q_{RB}$ since both the black body and the superconducting material optical properties do not vary between $T_l$ and $T_h$ which leads to zero rectification. When $T_h$ slightly surpasses $T_c$, the superconducting material becomes less reflecting in FB, \textit{i.e.} when $T_2 = T_h$, which explains the sharp increase of the exchanged RHF. Indeed, at $T_2 \gtrsim T_c$, in addition to the flux increase due to $\Delta T$ rise, FB RHF is strongly enhanced due to the non-linearity of the superconducting material TOP.

Now, the black body is replaced by another high temperature superconductor : YBa$_2$Cu$_3$O$_7$ with a phase transition at $T_c = 93$ K. Its reflectance data in the normal and superconducting states are given in Ref. \cite{Ginsberg1992}.
\begin{figure}[ht]
\begin{center}
\begin{subfigure}[b]{0.23\textwidth}
                \includegraphics[width=\textwidth]{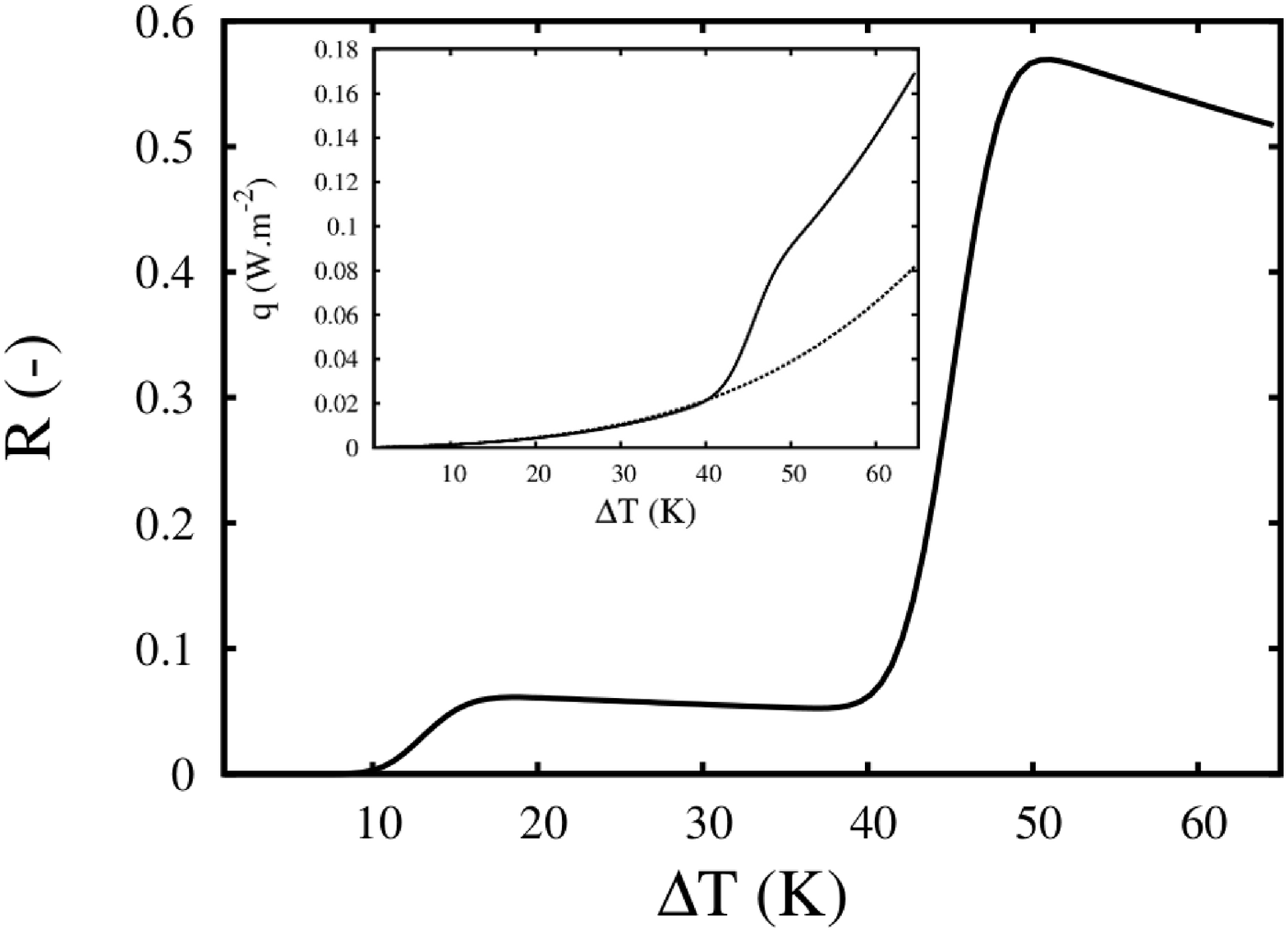}
                \caption{}
                \label{fig:QRvsDTdoubleA}
\end{subfigure}%
\quad
\begin{subfigure}[b]{0.23\textwidth}
                \includegraphics[width=\textwidth]{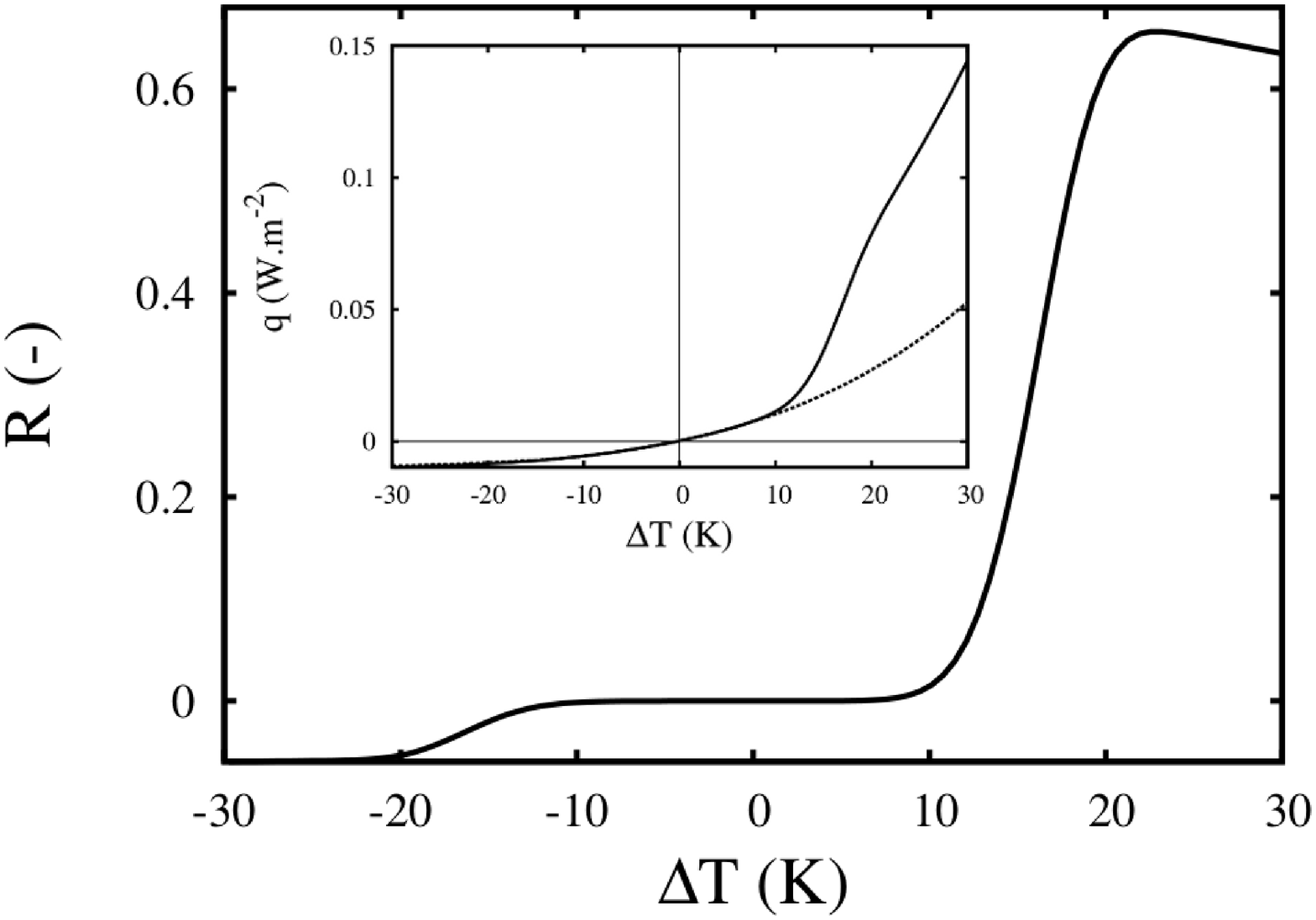}
                \caption{}
                \label{fig:QRvsDTdoubleB}
\end{subfigure}
\caption[]{(a) Rectification ratio for $T_l = 77$ K and $T_h = T_l + \Delta T$ as a function of the temperature difference $\Delta T$ ranging over $[1,65]$ K. (b) Rectification ratio for $T_l = (T_{c,1} + T_{c,2})/2$ K and $T_h = T_l + \Delta T$ as a function of the temperature difference $\Delta T$ ranging over $[-30,30]$ K. The inset of each figure shows the exchanged radiative heat flux density in FB (solid line) and RB (dashed line) as a function of $\Delta T$ for the corresponding configuration.}
\label{fig:QRvsDTdouble}
\end{center}
\end{figure}
We note $T_{c,1} = 93$ K and $T_{c,2} = 125$ K. As for the single superconductor device, $T_l$ is fixed at $T_l = 77$ K and $T_h$ is given by $T_h = T_l + \Delta T$ with $\Delta T$ ranging over $[1,65]$ K. Rectification ratio as a function of $\Delta T$ is presented in Fig. \ref{fig:QRvsDTdoubleA}. A maximal rectification of almost $R_{d,max} = 0.6$ is reached. This value is lower than the single superconductor rectifier maximal rectification ratio. However, we observe finite rectification over a broad temperature range starting at $T_h = T_{c,1} + \delta T$ with $\delta T \ll T_{c,1}$. After its maximal value, the rectification ratio of the double superconductors rectifier decreases almost three times faster than the single superconductor device ($\partial Ln(R_s) / \partial T = 2.5 \times 10^{-3}$ K$^{-1}$ versus $\partial Ln(R_d) / \partial T = 6.9 \times 10^{-3}$ K$^{-1}$). 
The net exchanged radiative heat flux density as a function of $\Delta T$ is presented in the inset of Fig. \ref{fig:QRvsDTdoubleA}. We observe that RB RHF density after maximal rectification ratio ($T \gtrsim T_{c,2}$) increases much faster than for a single superconductor rectifier. In the latter, two phenomena compete to increase and reduce exchanged RHF in RB : $\Delta T$ increase, and the superconductor emissivity drop in the superconducting state, respectively. In the present case, three phenomena are competing. The additional transition of YBa$_2$Cu$_3$O$_7$ from the superconducting to the normal state partially counter-balances the effect of Tl$_2$Ba$_2$CaCu$_2$O$_8$ transition and leads to a faster rise of $q_{RB}$, hence to a faster drop of the rectification ratio. 

Let us now consider a different situation where $T_l$ is fixed between the two critical temperatures, $T_l = (T_{c,1} + T_{c,2}) / 2 = 109$ K in this case, and $T_h$ is given by $T_h = T_l + \Delta T$ with $\Delta T$ ranging over $[-30,30]$ K. Corresponding fluxes in FB and RB scenarii as well as the rectification ratio as a function of $\Delta T$ are presented in Fig \ref{fig:QRvsDTdoubleB}. This figure tells a completely different story since, in this configuration, the two superconductors transitions do not have opposite effects anymore. First, we observe a negative rectification for $T_h < T_{c,1}$ and a positive rectification for $T_h > T_{c,2}$. Besides, the maximal rectification ratio increases to $R_{d,max} = 0.68$ when $T_h \gtrsim T_{c,2}$ while remaining lower than that of a single superconductor rectifier though. We present in the inset of Fig. \ref{fig:QRvsDTdoubleB} the net exchanged RHF density in forward and reverse bias. The exchanged RHF increases with $\Delta T$ but remains very small until a threshold $\Delta_t T = T_{c,2} - T_l$. The threshold effect is more spectacular for FB RHF. Not that the proposed device exhibits a very similar behavior to that of electric diodes usually observed in their I-V characteristics with the main difference that, in the present case, the current density is not zero for negative thermal potential differences.

We presented in this letter a theoretical concept of a thermal rectifier based on far-field thermal radiation between high temperature superconducting materials operating below $150$ K. We examined two configurations : a superconductor (Tl$_2$Ba$_2$CaCu$_2$O$_8$) exchanging with 1) a black body and 2) another superconductor, YBa$_2$Cu$_3$O$_7$ in this case. The first configuration proves to be more efficient. Indeed, it shows a maximal rectification ratio of $80 \%$ versus $70 \%$ for a temperature difference up to $60$ K and maintains a high rectification level in a larger operating temperature range. Finally, these performances may be enhanced by considering radiative heat transfer in the near-field especially since near-field effects onset distances at the considered temperatures are within reach of present experimental setups.
Presented results might be useful for energy conversion devices, efficient cryogenic radiative insulators engineering and thermal logical circuits development.

Authors gratefully acknowledge the support of the Agence Nationale de la Recherche through the Source-TPV Project No. ANR 2010 BLAN 0928 01. This work pertains to the French Government program “Investissements d’Avenir” (LABEX INTERACTIFS, reference ANR-11-LABX-0017-01). 
%
\end{document}